\documentclass[12pt]{article}
\usepackage{epsfig,latexsym,amssymb,amsmath}
\usepackage{color}
\usepackage{graphicx}
\begin{document}

\begin{center}
{\large \bf Thermodynamic Properties of Kehagias-Sfetsos Black Hole \& KS/CFT Correspondence}
\end{center}

\vskip 5mm

\begin{center}
{\Large{Parthapratim Pradhan\footnote{E-mail: pppradhan77@gmail.com}}}
\end{center}

\vskip  0.5 cm

{\centerline{\it Department of Physics}}
{\centerline{\it Hiralal Mazumdar Memorial College For Women }}
{\centerline{\it Dakshineswar, Kolkata-700035, India}}

\vskip 1cm

\begin{abstract}
We speculate on various thermodynamic features of the inner horizon~(${\mathcal H}^{-}$) and outer 
horizons~(${\mathcal H}^{+}$) of Kehagias-Sfetsos~(KS) black hole~(BH) in the background of 
Ho\v{r}ava Lifshitz gravity. We compute particularly the
\emph{area product, area sum, area minus and area division} of the BH horizons. We find that they all are 
\emph{not} showing universal behavior whereas the product is a universal quantity~
[Pradhan P., \textit{Phys. Lett. B}, {\bf 747} (2015) {64}]. 
Based on these relations, we derive the area bound  of all horizons. From the area bound we derive the 
entropy bound and  irreducible mass bound for all the horizons~(${\mathcal H}^{\pm}$). 
We also observe that the \emph{First law} of BH thermodynamics and \emph {Smarr-Gibbs-Duhem } relations do 
not hold for this BH. The underlying reason behind this failure due to the scale invariance of the coupling 
constant. Moreover, we compute the \emph{Cosmic-Censorship-Inequality} for this BH which gives the lower bound 
for the total mass of the spacetime and it is supported by cosmic cencorship conjecture. Finally, we discuss the 
KS/CFT~(Conformal Field Theory) correspondence via a thermodynamic procedure. 
\end{abstract}

\section{Introduction}

Recently the general relativity  community and the string theory community have become quite 
interested in examining the thermodynamic features of~${\mathcal H}^{-}$ and ${\mathcal H}^{+}$ 
~\cite{ah09,cgp11,cr12,sd12,pope14,pp14,ac79,mv13}. Of particular interest are relations that are 
independent of mass, so called ADM~(Arnowitt-Deser-Misner) mass, and then these relations are said 
to be ``universal'' in BH physics.  They are novel in the sense that they involve the thermodynamic 
quantities defined at multi-horizons, i.e. the Cauchy~(inner) horizon and event~(outer) horizons of 
the spherically symmetric charged, axisymmetric charged and axisymmetric non-charged BH. For example, 
let us consider first spherically symmetric charged BH, i.e., Reissner Nordstr{\o}m~(RN) BH, the 
mass-independent relation  for both the horizons~(${\mathcal H}^{\pm}$) becomes
\begin{eqnarray}
{\cal A}_{+} {\cal A}_{-} &=& (4\pi Q^2)^2 \,\,\, \mbox{or} \,\,\, {\cal S}_{+} {\cal S}_{-} = (\pi Q^2)^2
~.\label{pa}
\end{eqnarray}
For spinning non-charged BH,  i.e. for Kerr BH, the mass independent relations are
\begin{eqnarray}
{\cal A}_{+} {\cal A}_{-} &=& (8\pi J)^2 \,\,\, \mbox{or} \,\,\, {\cal S}_{+} {\cal S}_{-} = (2\pi J)^2 ~.\label{prK}
\end{eqnarray}

Finally, for charged spinning BH, these relations should read
$$
{\cal A}_{+} {\cal A}_{-} = (8\pi J)^2+(4\pi Q^2)^2
$$
or
\begin{eqnarray}
{\cal S}_{+} {\cal S}_{-} & =& (2\pi J)^2 + (\pi Q^2)^2  ~.\label{prKN}
\end{eqnarray}
Remarkably, all these thermodynamic relations are independent of the mass parameter therefore it should 
be treated as a universal quantity.

For the BPS~(Bogomol'ni-Prasad-Sommerfield) class of BHs, the area product formula~\cite{cgp11} of 
${\mathcal H}^{\pm}$ should be written as
$$
{\cal A}_{+} {\cal A}_{-} = (8\pi)^2 \left(\sqrt{N_{L}}+\sqrt{N_{R}}\right)\left(\sqrt{N_{L}}-\sqrt{N_{R}}\right)
$$
\begin{eqnarray}
= N , \,\, N\in {\mathbb{N}}, N_{L}\in {\mathbb{N}}, N_{R} \in {\mathbb{N}} ~.\label{ppl}
\end{eqnarray}
where the integers $N_{L}$ and $N_{R}$ should be defined as excitation numbers of the left- and right- 
moving sectors of a weakly-coupled 2D conformal field theory~(CFT). Resultantly, the entropy product 
formula of ${\mathcal H}^{\pm}$ becomes
$$
{\cal S}_{+} {\cal S}_{-} = (2\pi)^2 \left(\sqrt{N_{L}}+\sqrt{N_{R}}\right)\left(\sqrt{N_{L}}-\sqrt{N_{R}}\right)
$$
\begin{eqnarray}
= N , \,\, N\in {\mathbb{N}}, N_{L}\in {\mathbb{N}}, N_{R} \in {\mathbb{N}} ~.\label{ss}
\end{eqnarray}
This implies that the product of the entropy of ${\mathcal H}^{\pm}$ is an integer quantity~\cite{fl97}.

The product formulae that we would like to derive in this work, either area or entropy product of inner 
horizon and outer horizons could be used to determine whether the corresponding Bekenstein-Hawking 
entropy may be written as a Cardy formula, therefore providing some evidences for a CFT description 
of the corresponding microstates~\cite{cr12,sd12,chen12}. This boosts the study of the properties of 
the inner horizon thermodynamics in contrast with the outer horizon thermodynamics.

In our previous study~\cite{pp15}, we derived the surface area product, BH entropy product, surface 
temperature product, Komar energy product and specific-heat product for this BH. Besides the area or
entropy product it should be known what happens in case of \emph{area sum, area minus and area division}.  
For this reason we extend our study by computing area sum, entropy sum, temperature sum and specific- 
heat sum of all the horizons. We expect that the quantization area product formula that we have found 
from our previous investigation and from present study provides a strong indication that there exists 
an universal near-horizon structure for more general class of BHs. This indicates the possibility that 
the microscopic degrees of freedom may admit a dual field theoretic explanation that generalizes the 
2D CFT duals. 

Thus in this Letter, we wish to examine various thermodynamic features~(besides the area or entropy product) 
of Kehagias-Sfetsos BH~\cite{ks09} in Ho\v{r}ava Lifshitz gravity~\cite{ph9a,ph9b,ph9c}. We have considered 
both the inner horizon and outer horizons to  further understanding the microscopic nature of BH entropy both 
interior and exterior. Moreover using these relations, we derive the area bound of all horizons. From area 
bound we derive entropy bound and irreducible mass bound for both the horizons. 

One aspect that has not been studied previously  is so called the \emph{Cosmic-Censorship-Inequality} or the
\emph{Cosmic Censorship Bound}~\cite{gibb05}. It should require the cosmic-censorship hypothesis~\cite{rp73} 
(See~\cite{bray,bray1,jang,rg,gibb99}) and which is an important inequality in general relativity which relates 
the total mass of the spacetime in terms of the ${\mathcal H}^{+}$ area, and for Schwarzschild BH it should be 
minimum i.e.
\begin{eqnarray}
{\cal M}  &\geq&  \sqrt{\frac{{\cal A}_{+}}{16\pi}} ~.\label{spi}
\end{eqnarray}
This brilliant idea was first given by Penrose in 1973~\cite{rp73}~\footnote{ This beautiful argument can translate 
into a very interesting mathematical inequality in Riemannian geometry which is so called 
\emph{Riemannian Penrose Inequality}. It was first examined and proved by  Huisken et al.~\cite{hui}. 
This inequality has an important application in 
gravitational collapse and using Cauchy data it could be solved the Einstein equations. Finally, it has another interesting 
application to solve the Yamabe problem~\cite{bray}. It should be noted that \emph{Riemannian Penrose Inequality} satisfied 
the \emph{Riemannian positive mass theorem}~\cite{yau}. }.

This inequality has an important implication in BH physics that it indicates the lower bound on the energy for 
a time-symmetric initial Cauchy data set which satisfies the Einstein equations, and which 
has also  satisfied the dominant energy condition and which has  no naked singularities.

The structure of the paper is as follows. In the second section, we shall describe various thermodynamic features 
of Kehagias-Sfetsos BH in Ho\v{r}ava Lifshitz gravity, we also calculate the different thermodynamic bound 
in different subsections. In the third section, we discuss the KS/CFT correspondence using thermodynamic 
procedure. Finally, we conclude our discussions in the last section.

\section{Thermodynamic Properties of KS BH in Ho\v{r}ava Lifshitz gravity}
At Lifshitz point, Ho\v{r}ava ~\cite{ph9a,ph9b,ph9c} has given a beautiful theory for general relativity which is 
renormalizable and UV complete. It can be reduced to Einstein's general relativity at large 
 scales for the value of dynamical coupling constant $\lambda=1$. We have not mentioned here in detail the ADM 
 formalism  because it has been already mentioned in~ \cite{pp15}. Since we are interested in this work to study 
 the thermodynamic properties of Kehagias-Sfetsos~(KS) BH ~\cite{ks09} in Ho\v{r}ava Lifshitz ~(HL) gravity, thus 
 the  metric of KS BH ~\cite{lmp09,ym09,ks09,mk10,cc9a,cc9b} is given by
\begin{eqnarray}
ds^2 &= & -{\cal F}(r) dt^2 + \frac{dr^2}{{\cal F}(r)} +r^2(d\theta^2+ \sin^2\theta d\phi^2).~\label{hl}
\end{eqnarray}
where,
\begin{eqnarray}
{\cal F}(r) &=&  1-\sqrt{4 \mathcal{M} \omega r+\omega^2 r^4} +\omega r^2,\label{hl1}
\end{eqnarray}
and $\mathcal{M}$ is an integration constant derived from equations of motion of KS action. This constant is 
treated as ``mass'' parameter in HL gravity. For $r\gg (\frac{{\cal M}}{\omega})^{\frac{1}{3}}$, one obtains 
the result of a Schwarzschild BH.

The BH horizons occur at ${\cal F}(r)=0$:
\begin{eqnarray} 
r_{\pm}={\cal M} \pm \sqrt{\mathcal{M}^2 -\frac{1}{2\omega}} ~.\label{hl2}
\end{eqnarray}
where $r_{+}$ is the event horizon and  $r_{-}$ is the Cauchy horizon respectively. As long as
\begin{eqnarray}
{\cal M}^2 -\frac{1}{2\omega} \geq 0 ~.\label{ineq}
\end{eqnarray}
then the KS metric  describes a BH, otherwise it has a naked  singularity. When ${\cal M}^2 -\frac{1}{2\omega}=0$, 
we find the extremal KS BH.

The product and sum  of horizon radii become
\begin{equation}
r_+ r_- = \frac{1}{2\omega}\,\,\, \mbox{and} \,\, r_{+}+ r_{-} = 2{\cal M} ~.\label{hl3}
\end{equation}

The area~\cite{pp15} of this BH is given by
\begin{eqnarray}
{\cal A}_{\pm} &=& 4\pi\left(2\mathcal{M}r_{\pm}-\frac{1}{2\omega} \right)
\end{eqnarray}
Their product~\cite{pp15} and sum yield
\begin{eqnarray}
{\cal A}_{+}{\cal A}_{-} &=&  \frac{4\pi^2}{\omega^2} \,\,\, \mbox{and} \,\,\, {\cal A}_{+}+{\cal A}_{-} =
4\pi\left(4{\cal M}^2-\frac{1}{\omega} \right)  ~. \label{psks}
\end{eqnarray}
It is remarkable that the area product of KS BH is independent of mass but the area sum is not independent
of the mass parameter. 

For completeness, we further compute the area minus and area division:
\begin{eqnarray}
{\cal A}_{\pm}- {\cal A}_{\mp} &=& \pm 16 \pi {\cal M} \sqrt{{\cal M}^2-\frac{1}{2\omega}}  ~.\label{area-}
\end{eqnarray}
and  
\begin{eqnarray}
\frac{{\cal A}_{+}}{{\cal A}_{-}} &=& \frac{r_{+}^2}{r_{-}^2} ~.\label{aread}
\end{eqnarray}

Again, the sum of area inverse is found to be 
\begin{eqnarray}
\frac{1}{{\cal A}_{+}}+\frac{1}{{\cal A}_{-}} &=& \frac{\omega^2}{\pi} \left(4{\cal M}^2-\frac{1}{\omega} \right)
 ~.\label{arid}
\end{eqnarray}
and the minus of area inverse is computed to be 
\begin{eqnarray}
\frac{1}{{\cal A}_{\pm}}-\frac{1}{{\cal A}_{\mp}} &=& \mp \frac{4\omega^2{\cal M}}{\pi} 
\sqrt{{\cal M}^2-\frac{1}{2\omega}} ~.\label{areid-}
\end{eqnarray}
It indicates that they are all mass dependent relations.

Likewise, the entropy product ~\cite{pp15} and entropy sum of ${\cal H}^{\pm}$ become
\begin{eqnarray}
{\cal S}_{-} {\cal S}_{+}  &=& \frac{\pi^2}{4\omega^2} \,\,\, \mbox{and} \,\,\,
{\cal S}_{-}+{\cal S}_{+}  =  \pi\left(4{\cal M}^2-\frac{1}{\omega} \right) 
~.\label{etpks}
\end{eqnarray}

For record, we also compute the entropy minus of ${\cal H}^\pm$ as 
\begin{eqnarray}
{\cal S}_{\pm}- {\cal S}_{\mp} &=&  \pm 4 \pi {\cal M} \sqrt{{\cal M}^2-\frac{1}{2\omega}}  ~.\label{entr-}
\end{eqnarray}
and  the entropy division of ${\cal H}^{\pm}$ as
\begin{eqnarray}
\frac{{\cal S}_{+}}{{\cal S}_{-}} &=& \frac{r_{+}^2}{r_{-}^2}  ~.\label{entrd}
\end{eqnarray}
Again, the sum of entropy inverse is found to be 
\begin{eqnarray}
\frac{1}{{\cal S}_{+}}+\frac{1}{{\cal S}_{-}} &=& \frac{4\omega^2}{\pi} \left(4{\cal M}^2-\frac{1}{\omega} \right)
~.\label{etpid}
\end{eqnarray}
and the minus of entropy inverse is 
\begin{eqnarray}
\frac{1}{{\cal S}_{\pm}}-\frac{1}{{\cal S}_{\mp}} &=& \mp \frac{16\omega^2{\cal M}}{\pi} 
\sqrt{{\cal M}^2-\frac{1}{2\omega}}
~.\label{etpid-}
\end{eqnarray}
The Hawking ~\cite{bcw73} temperature on ${\cal H}^{\pm}$ reads 
\begin{eqnarray}
T_{\pm} &=& \frac{\omega (r_{\pm}- {\cal M})}{2\pi(1+\omega r_{\pm}^2)}.~\label{M9}
\end{eqnarray}
Their product~ \cite{pp15} and sum yield
$$
T_{+} T_{-} = \frac{\omega \left(1-2{\cal M}^2\omega \right)}{2 \pi^2 (1+16{\cal M}^2\omega)} 
$$
and
\begin{eqnarray} 
T_{+} +T_{-} = \frac{4\omega {\cal M} \left(1-2{\cal M}^2\omega \right)}
{\pi \left(1+16{\cal M}^2\omega \right)}  .~\label{M12}
\end{eqnarray}
It may be noted that surface temperature product and sum both depends on mass, thus they are 
not universal in nature. It is shown that for KS BH
\begin{eqnarray}
T_{+}{\cal S}_{+}+T_{-}{\cal S}_{-} &=& \frac{8 \omega {\cal M} \sqrt{{\cal M}^2-\frac{1}{2\omega}}}
{1+16 \omega {\cal M}^2} .~\label{ts}
\end{eqnarray}
In general, this relation is for RN BH or Kerr BH \cite{ok92}
\begin{eqnarray}
T_{+}{\cal S}_{+}+ T_{-}{\cal S}_{-} &=& 0 ~.\label{exrk1}
\end{eqnarray}
It is in fact a mass independent~(universal) relation and implies that $T_{+}{\cal S}_{+}=- T_{-}{\cal S}_{-}$
should be taken as a criterion whether there is a 2D CFT dual for the BHs in the Einstein gravity and other 
diffeomorphism gravity theories~\cite{chen12,xu}. This universal relation also indicates that the left and right 
central charges are equal i.e., $c_{L}=c_{R}=12 J$ which is holographically dual to 2D CFT~\cite{ac10}.

But for KS BH, it follows from Eq. (\ref{ts}) that it is mass dependent and it does not vanishes as in Eq. (\ref{exrk1}) 
that means the central charges of the left moving sectors and right moving sectors are \emph{not} equal. 
This is an interesting observation for KS BH in HL gravity wherea Einstein gravity does not possesses such type of 
features. It  is also  interesting to mentioned that except the area~(or entropy) product and irreducible mass product 
all the thermodynamic relations of KS BH are \emph{mass dependent}.

\subsection{Smarr Formula for HL BH on $\mathcal{H}^{\pm}$}
Smarr \cite{ls73} had first derived  the  mass parameter can be expressed as in terms  of area, angular momentum and
charge for Kerr-Newman BH. On the otherhand,  Hawking \cite{bcw73} has been speculated that the BH area
always increases. Therefore the BH  area is indeed a constant quantity over the ${\cal H}^{\pm}$.
Analogously, the area of both the horizons for KS BH in HL gravity is given by
\begin{eqnarray}
\mathcal{A}_{\pm}=4 \pi \left[ 2\mathcal{M}^2 -\frac{1}{2\omega} \pm 2\mathcal{M} \sqrt{\mathcal{M}^2-\frac{1}{2\omega}} \right]
\label{M14}
\end{eqnarray}

Alternatively, the mass parameter could be expressed as, in terms of horizons~(${\cal H}^{\pm}$),
\begin{eqnarray}
\mathcal{M}^2 &=& \frac{\mathcal{A}_{\pm}}{16 \pi}+\frac{\pi }{4\omega^2 \mathcal{A}_{\pm}}+\frac{1}{4\omega}.~\label{M16} 
\end{eqnarray}
Form the above relation we can easily derived the \emph{Cosmic-Censorship-Inequality} for KS BH
\begin{eqnarray}
\mathcal{M} &\geq& \sqrt{\frac{\mathcal{A}_{\pm}}{16 \pi}+\frac{\pi }{4\omega^2 \mathcal{A}_{\pm}}+\frac{1}{4\omega}}.
~\label{rmp16} 
\end{eqnarray}
Actually, Penrose derived it for ${\cal H}^{+}$ only. We here suggest this inequality is valid for ${\cal H}^{-}$ also.

After differentiation, we get the mass differential as
\begin{eqnarray}
d\mathcal{M}=\mathcal{T_{\pm}} d\mathcal{A}_{\pm} +\Phi_{\pm}^{\omega}d\omega ~\label{M17}
\end{eqnarray}
where,
\begin{eqnarray}
\mathcal{T}_{\pm} &=& \text{Effective surface tension for horizons}\nonumber\\
                  &=& \frac{1}{\mathcal{M}}\Big(\frac{1}{32\pi}-\frac{\pi}{8 \omega^2 \mathcal{A}_{\pm}^2}\Big)=
                       \frac{\partial {\cal M}}{\partial {\cal A}_{\pm}}.\\
\Phi_{\pm}^{\omega} &=& \text{Effective potential for horizons due to } \omega \nonumber\\
           &=& - \frac{1}{\mathcal{M}}\Big(\frac{\pi}{4\omega^3 \mathcal{A}_{\pm}}+\frac{1}{8\omega^2}\Big)\\
           &=& -\frac{1}{4\omega^2 r_{\pm}}
           = \frac{\partial {\cal M}}{\partial \omega}. ~\label{M18}
\end{eqnarray}
It is well known that for spherically symmetric RN BH, the Smarr-Gibbs-Duhem relation is satisfied by the 
following condition:
\begin{eqnarray}
\frac{\cal M}{2}-T_{\pm}{\cal S}_{\pm}-\frac{\Phi_{\pm}}{2}Q &=& 0 .~\label{sgd}
\end{eqnarray}
where the symbols are used as usual for RN BH. But for KS BH this relation is 
\begin{eqnarray}
\frac{\cal M}{2}-T_{\pm}{\cal S}_{\pm}-\frac{\Phi_{\pm}^{\omega}}{2}\omega &=& 
\frac{2+5\omega r_{\pm}^2}{8\omega r_{\pm}(1+\omega r_{\pm}^2)} \neq 0 .~\label{sgd1}
\end{eqnarray}
It indicates that the Smarr-Gibbs-Duhem relation do not satisfied for KS BH in HL gravity. Followed by 
the first law of BH thermodynamics which is also satisfied for this BH. The reason should be due to the 
scale invariance of the coupling constant $\omega$. This observation is essential here because we have 
not seen such a type of discussion in the literature regarding the KS BH in HL gravity.

It should be emphasized that when we add the AdS term to this BH then the both first law of thermodynamics and 
Smarr-Gibbs-Duhem relations have satisfied which has been explicitly examined in~\cite{brena}. Where the author 
derived the generalized Smarr relation in AdS space which has include a pressure-volume term and the thermodynamic 
mass, ADM mass, Brown-York mass and Holland-Ishibashi-Marolf mass could also be defined.  But it is interesting 
to note that with out pressure-volume term the first law and Smarr relation do not satisfied at all. This is one of 
the key results of our work.

\subsection{Area Bound of KS BH for ${\cal H}^{\pm}$}
Using the above thermodynamic relations, we are now able to derive the entropy bound of both the horizons.
Using the ineqality equation~(\ref{ineq}) one can obtain ${\cal M}^2 \geq \frac{1}{2\omega}$. Since 
$r_{+} \geq r_{-}$, one can get ${\cal A}_{+} \geq {\cal A}_{-} \geq 0$. Then the area product gives
\begin{eqnarray}
{\cal A}_{+}  \geq  \sqrt{{\cal A}_{+} {\cal A}_{-}}=\frac{2\pi}{\omega} \geq {\cal A}_{-} ~.\label{ieqa1}
\end{eqnarray}
and the area sum gives
$$
4\pi\left(4{\cal M}^2-\frac{1}{\omega} \right) ={\cal A}_{+}+ {\cal A}_{-}  \geq 
$$
\begin{eqnarray}
{\cal A}_{+} \geq 
\frac{{\cal A}_{+}+ {\cal A}_{-}}{2}= 2\pi \left(4{\cal M}^2-\frac{1}{\omega} \right) 
 ~.\label{inqa2}
\end{eqnarray}
Thus the area bound for ${\cal H}^{+}$ satisfies
\begin{eqnarray}
 2\pi \left(4{\cal M}^2-\frac{1}{\omega} \right)  \leq {\cal A}_{+} \leq 4\pi \left(4{\cal M}^2-\frac{1}{\omega} \right)
 ~.\label{inqa3}
\end{eqnarray}
and  the area bound for  ${\cal H}^{-}$ satisfies
\begin{eqnarray}
 0 \leq {\cal A}_{-} \leq \frac{2\pi}{\omega} ~.\label{inqa4}
\end{eqnarray}

\subsection{Entropy Bound for ${\cal H}^{\pm}$}
Analogously, as $r_{+} \geq r_{-}$, one can get ${\cal S}_{+} \geq {\cal S}_{-} \geq 0$. 
Then the entropy product gives
\begin{eqnarray}
{\cal S}_{+}  \geq  \sqrt{{\cal S}_{+} {\cal S}_{-}}=\frac{\pi}{2\omega} \geq {\cal S}_{-} ~.\label{si1}
\end{eqnarray}
and the entropy sum gives
$$
\pi\left(4{\cal M}^2-\frac{1}{\omega} \right) ={\cal S}_{+}+ {\cal S}_{-} \geq {\cal S}_{+} \geq
$$
\begin{eqnarray} 
\frac{{\cal S}_{+}+ {\cal S}_{-}}{2}= \pi \left(2{\cal M}^2-\frac{1}{2\omega} \right) 
 ~.\label{si2}
\end{eqnarray}
Thus the entropy bound for  ${\cal H}^{+}$ satisfies
\begin{eqnarray}
 \pi \left(2{\cal M}^2-\frac{1}{2\omega} \right)  \leq {\cal S}_{+} \leq \pi \left(4{\cal M}^2-\frac{1}{\omega}\right)
 ~.\label{si3}
\end{eqnarray}
and  the entropy bound for ${\cal H}^{-}$
\begin{eqnarray}
 0 \leq {\cal S}_{-} \leq \frac{\pi}{2\omega} ~.\label{si4}
\end{eqnarray}

\subsection{ Irreducible mass bound for ${\cal H}^{\pm}$}
Christodoulou~\cite{cd70} had given a relation between surface area of the ${\cal H}^{+}$ 
and irreducible mass,  which can be written as 
\begin{eqnarray}
{\cal M}_{\text{irr}, +}^{2} &=& \frac{{\cal A}_{+}}{16\pi}=\frac{{\cal S}_{+}}{4\pi}
~. \label{irrm}
\end{eqnarray}
It is now well known that this relation is valid for CH too. That means
\begin{eqnarray}
{\cal M}_{\text{irr}, -}^{2} &=& \frac{{\cal A}_{-}}{16\pi}=\frac{{\cal S}_{-}}{4\pi}
~. \label{irrm1}
\end{eqnarray}
Now the the product and sum  of the irreducible mass for both the horizons are
$$
{\cal M}_{\text{irr}, +} {\cal M}_{\text{irr},-} = \frac{1}{8\omega}
$$
and
\begin{eqnarray}
{\cal M}_{\text{irr}, +}^{2} + {\cal M}_{\text{irr},-}^{2} = {\cal M}^2-\frac{1}{4\omega} 
.~\label{M22}
\end{eqnarray}
From the area bound, we get the irreducible mass bound for KS BH
\begin{eqnarray}
\frac{\sqrt{4{\cal M}^2-\frac{1}{\omega}}}{2\sqrt{2}} \leq {\cal M}_{irr, +} \leq  
\frac{\sqrt{4{\cal M}^2-\frac{1}{\omega}}}{2}  ~.\label{inqa5}
\end{eqnarray}
and 
\begin{eqnarray}
0 \leq {\cal M}_{irr,-} \leq \sqrt{\frac{1}{8\omega}}  ~.\label{inqa6}
\end{eqnarray}
Eq. \ref{inqa5} is nothing but the Penrose inequality, which is the first geometric inequality for 
BHs~\cite{rg}.

\subsection{Temperature Bound for ${\cal H}^{\pm}$}
In BH thermodynamics, temperature is an important parameter. So there must exist temperature bound
relation on the horizons. As is when  $r_{+} \geq r_{-}$,  one must obtain $T_{+} \geq T_{-} \geq 0$.
Then the temperature product gives
\begin{eqnarray}
T_{+}  \geq  \sqrt{T_{+} T_{-}}=
\sqrt{\frac{\omega \left(1-2{\cal M}^2\omega \right)}{2 \pi^2 (1+16{\cal M}^2\omega)} } 
\geq T_{-} ~.\label{iqt1}
\end{eqnarray}
and the temperature sum gives
$$
\frac{4\omega {\cal M} \left(1-2{\cal M}^2\omega \right)}
{\pi \left(1+16{\cal M}^2\omega \right)} =T_{+}+ T_{-} \geq T_{+} \geq 
$$
\begin{eqnarray}
\frac{T_{+}+ T_{-}}{2}= \frac{2\omega {\cal M} \left(1-2{\cal M}^2\omega \right)}
{\pi \left(1+16{\cal M}^2\omega \right)} 
 ~.\label{iqt2}
\end{eqnarray}
Thus, the temperature bound for  ${\cal H}^{+}$
\begin{eqnarray}
 \frac{2\omega {\cal M} \left(1-2{\cal M}^2\omega \right)}
{\pi \left(1+16{\cal M}^2\omega \right)} \leq T_{+} \leq \frac{4\omega {\cal M} \left(1-2{\cal M}^2\omega \right)}
{\pi \left(1+16{\cal M}^2\omega \right)}
 ~.\label{iqt3}
\end{eqnarray}
and  the temperature bound for  ${\cal H}^{-}$
\begin{eqnarray}
0 \leq T_{-} \leq \sqrt{\frac{\omega \left(1-2{\cal M}^2\omega \right)}{2 \pi^2 (1+16{\cal M}^2\omega)} } 
 ~.\label{iqt4}
\end{eqnarray}

\subsection{Bound on heat capacity $C_{\pm}$ for ${\cal H}^{\pm}$}
In BH thermodynamics, the specific heat can be defined as 
\begin{eqnarray}
C_{\pm} &=& \frac{\partial{\cal M}}{\partial T_{\pm}} .~\label{c1}
\end{eqnarray}
which is an important parameter to determine the thermodynamic properties in BH physics. In our previous
work\cite{pp15}, we derived in detail the expression for specific heat for both the horizons.
It is given by
\begin{eqnarray}
C_{\pm} &=& \frac{2 \pi  }{\omega} \frac{(2\omega r_{\pm}^2-1)\big(1+\omega r_{\pm}^2\big)^2}
{1+5\omega r_{\pm}^2-2\omega^2 r_{\pm}^4}  .~\label{c5}
\end{eqnarray}
Their product \cite{pp15} and sum  on ${\cal H}^{\pm}$ yields
\begin{eqnarray}
C_{+} C_{-} &=& \frac{\pi^2}{2\omega^2}\frac{\left(1-2{\cal M}^2\omega \right)\left(1+16{\cal M}^2
\omega \right)^2}{\left(2+13 \omega {\cal M}^2-16 \omega^2 {\cal M}^4\right)}  .~\label{c6}
\end{eqnarray}
and 
$$
C_{+}+C_{-} =
$$
\begin{eqnarray}
\frac{\pi}{\omega^2}\frac{\left(128 \omega^4 {\cal M}^6 +8\omega^3 {\cal M}^4 -42\omega^2 {\cal M}^2
+4\omega {\cal M}^2 +2\omega -1\right)}{\left(2+13 \omega {\cal M}^2-16 \omega^2 {\cal M}^4\right)}  .~\label{c7}
\end{eqnarray}
Using ${\cal M}^2 \geq \frac{1}{2\omega}$ with the product of heat capacity  and the sum of heat capacity, we get 
the bound on heat capacity for both the horizons.
For  ${\cal H}^{+}$
$$
\frac{\pi}{2\omega^2}\frac{\left(128 \omega^4 {\cal M}^6 +8\omega^3 {\cal M}^4 -42\omega^2 {\cal M}^2
+4\omega {\cal M}^2 +2\omega -1\right)}{\left(2+13 \omega {\cal M}^2-16 \omega^2 {\cal M}^4\right)} 
$$
$$
\leq C_{+} \leq
$$
\begin{eqnarray} 
\frac{\pi}{\omega^2}\frac{\left(128 \omega^4 {\cal M}^6 +8\omega^3 {\cal M}^4 -42\omega^2 {\cal M}^2
+4\omega {\cal M}^2 +2\omega -1\right)}{\left(2+13 \omega {\cal M}^2-16 \omega^2 {\cal M}^4\right)} 
 ~.\label{c8}
\end{eqnarray}
and for ${\cal H}^{-}$
\begin{eqnarray}
 0 \leq C_{-} \leq \sqrt{\frac{\pi}{\omega^2}\frac{\left(128 \omega^4 {\cal M}^6 +8\omega^3 {\cal M}^4 -42\omega^2 {\cal M}^2
+4\omega {\cal M}^2 +2\omega -1\right)}{\left(2+13 \omega {\cal M}^2-16 \omega^2 {\cal M}^4\right)}} ~.\label{c9}
\end{eqnarray}
It should be mentioned that all the above thermodynamic formulae might be suggested the possibility of an 
explanation for the microscopic nature of such BHs in terms of a field theory in more than two dimensions.

\section{KS/CFT Correspondence}
In this section, we would like to prove that the central charges $c_{R}$ and $c_{L}$ of the right and left moving sectors 
of the dual CFT in KS/CFT correspondence are \emph{not} same. To do this we should calculate the thermodynamic parameters 
in left moving sectors and right moving sectors by  using the definitions of 
$\beta_{R,L}=\beta_{+}\pm \beta_{-}$, $\beta_{\pm}=\frac{1}{T_{\pm}}$, 
$\Phi_{R,L}^{\omega}=\frac{\beta_{+}\Phi_{+}^{\omega}\pm \beta_{-}\Phi_{-}^{\omega}}{2\beta_{R,L}}$ and 
$S_{R,L}=\frac{(S_{+}\mp S_{-})}{2}$~\cite{cv96,cv97,cvf97}. 
Now we could easily derive the temperature and entropy for left moving sectors and right moving sectors as
\begin{eqnarray}
T_{L} &=& \frac{1}{8\pi {\cal M}}, \,\,\, 
T_{R} = \frac{\omega}{\pi} \frac {\sqrt{{\cal M}^2-\frac{1}{2\omega}}} {\left(1+4 \omega {\cal M}^2\right)}  \nonumber\\
S_{L} &=& \frac{\pi}{2} \left(4{\cal M}^2-\frac{1}{\omega} \right) , \,\,\, 
S_{R} = 2\pi {\cal M} \sqrt{{\cal M}^2-\frac{1}{2\omega}}    \nonumber\\
\Phi_{L}^{\omega} &=& \frac{1}{64 {\cal M}\omega^2 },  \,\,\, \Phi_{R}^{\omega} =-\frac{3{\cal M}}{4\omega \left(1+4 \omega {\cal M}^2\right)}
~.\label{ks10}
\end{eqnarray}
The first law of thermodynamics could be rewritten as in terms of right and left moving sectors
of dual CFT
\begin{eqnarray}
\frac{dM}{2} &=& T_{R} dS_{R}+\Phi_{R}^{\omega} d\omega    ~.\label{ks8} \\
             &=& T_{L} dS_{L}+\Phi_{L}^{\omega} d\omega    ~.\label{ks9}
\end{eqnarray}
Using Eq. (\ref{ks8}) \& Eq. (\ref{ks9}), one could determine the first law of thermodynamics for 
left moving sectors and right moving sectors of dual CFT
\begin{eqnarray}
d\omega &=& \frac{T_{L}}{\Phi_{R}^{\omega}-\Phi_{L}^{\omega}} dS_{L}- \frac{T_{R}}{\Phi_{R}^{\omega}-\Phi_{L}^{\omega}} 
dS_{R} ~.\label{ks11}
\end{eqnarray}
Using above Eq. (\ref{ks11}), one can find the  dimensionless temperature of the left and right
moving sectors of the dual CFT correspondence  
\begin{eqnarray}
T_{L}^{\omega} &=& \frac{T_{L}}{\Phi_{R}^{\omega}-\Phi_{L}^{\omega}}, \,\, T_{R}^{\omega} = 
\frac{T_{R}}{\Phi_{R}^{\omega}-\Phi_{L}^{\omega}}  ~.\label{ks12}
\end{eqnarray}
For KS BH, the values are
\begin{eqnarray}
T_{L}^{\omega} &=& -\frac{8\omega^2}{\pi} \frac{\left(1+4 \omega {\cal M}^2 \right)}{\left(1+52 \omega {\cal M}^2 \right)}
~.\label{ks13}
\end{eqnarray}
\&
\begin{eqnarray}
T_{R}^{\omega} &=& -\frac{64 \omega^3 {\cal M}}{\pi} 
\frac {\sqrt{{\cal M}^2-\frac{1}{2\omega}}} {\left(1+52\omega {\cal M}^2\right)}    ~.\label{ks14}
\end{eqnarray}
Now we compute the central charges~\cite{chen12,guica} in left and right moving sectors of the KS/CFT 
correspondence using the Cardy formula 
\begin{eqnarray}
S_{L}^{\omega} &=& \frac{\pi^2}{3}c_{L}^{\omega}T_{L}^{\omega},\,\,\, S_{R}^{\omega} = 
\frac{\pi^2}{3}c_{R}^{\omega}T_{R}^{\omega}~.\label{ks14.5}
\end{eqnarray}
Therefore the central charges of dual CFT are 
\begin{eqnarray}
c_{L}^{\omega} &=& \frac{3\left(1-4\omega {\cal M}^2\right) \left(1+52\omega {\cal M}^2\right)}
{16\omega^3 \left(1+4\omega {\cal M}^2\right)  }  ~.\label{ks15}
\end{eqnarray}
\&
\begin{eqnarray}
c_{R}^{\omega} &=& - \frac{3 \left(1+52\omega {\cal M}^2\right)} {32 \omega^3}  ~.\label{ks16}
\end{eqnarray}
From the above calculation we prove that 
\begin{eqnarray}
c_{L} & \neq & c_{R}  ~.\label{ks17}
\end{eqnarray}
Now we could see what happens in the extreme limit?
\begin{eqnarray}
T_{L} &=& \frac{\sqrt{2\omega}}{8\pi }, \,\,\,  T_{R} = 0 \nonumber\\
S_{L} &=&  \frac{\pi}{2\omega}  , \,\,\, 
S_{R} = 0 \nonumber\\
\Phi_{L} &=& \frac{\sqrt{2\omega}}{64\omega^2},  \,\,\, \Phi_{R} =-\frac{1}{4\omega\sqrt{2\omega}} ~.\label{ks18}
\end{eqnarray}
Analogously the central charges  are 
\begin{eqnarray}
c_{L}^{\omega} &=& -\frac{27}{16 \omega^3}  ~.\label{ks19}
\end{eqnarray}
\&
\begin{eqnarray}
c_{R}^{\omega} &=& - \frac{81}{32\omega^3}  ~.\label{ks20}
\end{eqnarray}
Thus, the ratio of $c_{L}$ and $c_{R}$ is given by 
\begin{eqnarray}
 \frac{c_{L}}{c_{R}} &=& \frac{2}{3} ~.\label{ks21}
\end{eqnarray}
As we have said earlier in Eq. \ref{exrk1} and in Eq. \ref{ks17}, the central charges are not equal thus we could 
not find macroscopic Bekenstein-Hawking entropy of extreme KS BH. This is an interesting result of KS BH in HL gravity.

\section{Discussion:}
In order to understand the BH entropy~(both outer as well as inner) at the microscopic level, we
studied thermodynamic 
properties of KS BH in HL gravity. We computed various thermodynamic formula for this BH. 
We speculated that area sum, area minus and area division are mass dependent quantities, 
whereas the product \cite{pp15} is a mass independent quantity. 

Based on these relations, we computed area bound, entropy bound, irreducible mass bound, 
temperature bound and specific-heat bound. The upper area bound of outer horizon is actually
the Penrose-like inequality in BH mechanics. Due to the scale invariance of the coupling constant 
parameter $\omega$, we showed that the First law of BH thermodynamics and Smarr-Gibbs-Duhem  
relations do not satisfied for this BH. Finally, we derived the Cosmic-Censorship-Inequality for this 
BH which has an important implications in Cosmic-Censorship-Conjecture. 

We proved that the central charges  of KS BH in HL gravity are not equal and do not produce the 
macroscopic Bekenstein-Hawking entropy of the extreme KS BH which is a drawback of HL gravity.

In conclusion, these thermodynamic product formulae suggests further evidence for the crucial role of
both inner horizon and outer horizon for understanding the microscopic nature of BH entropy 
~(both interior and exterior) which is the prime aim in quantum gravity.

\section*{Acknowledgements}
The author is grateful to the authority of Inter-University Centre for Astronomy and 
Astrophysics~(IUCAA), Pune for warm hospitality during a ``Refresher course in Astronomy and 
Astrophysics'' and where the most of the work was done. The author also would like to thank 
Dr. W. G. Brenna for useful correspondence.


\begin{thebibliography}{99}
\bibitem{ah09} M. Ansorg \& J. Hennig, 
\textit{Phys. Rev. Lett.} {\bf 102}, {221102} (2009).

\bibitem{cgp11} M. Cveti\v{c} et al., 
\textit{  Phys. Rev. Lett.} {\bf 106}, {121301} (2011).

\bibitem{cr12} A. Castro \& M. J. Rodriguez, 
\textit{ Phys. Rev.} {\bf D 86}, {024008} (2012).

\bibitem{sd12} S. Detournay, 
\textit{Phys. Rev. Lett. } {\bf 109}, {031101} (2012).

\bibitem{pope14} M. Cveti\v{c} et al., \textit{Phys. Rev.} {\bf D 88}, {044046} (2013).

\bibitem{pp14} P. Pradhan, 
\textit{ Euro. Phys. J. C} {74},  {2887} (2014).

\bibitem{pp15} P. Pradhan,
\textit{Phys. Lett. B} {\bf 747}, {64} (2015).

\bibitem{ac79} A. Curir,  \textit{ Nuovo Cimento} {\bf 51B }, {262} (1979).

\bibitem{mv13} M. Visser, 
\textit{Phys. Rev.} {\bf D 88},  {044014} (2013).

\bibitem{fl97} F. Larsen, 
\textit{Phys. Rev.} {\bf D 56} {1005} (1997).

\bibitem{chen12} B.~Chen et al., 
\textit{ JHEP}  {\bf 017}, 1211 (2012).


\bibitem{ks09} A. Kehagias and K. Sfetsos, \textit{Phys. Lett. } {B 678}, {123} (2009).

\bibitem{ph9a} P. Horava, \textit{Phys. Rev. Lett. } {\bf 102}, {161301} (2009).

\bibitem{ph9b} P. Horava, \textit{Phys. Rev. } {\bf D 79}, {084008} (2009).

\bibitem{ph9c} P. Horava, \textit{JHEP} {\bf 03}, {020} (2009).

\bibitem{rp73} R. Penrose, \textit{ Ann. N. Y. Acad. Sci. } {\bf 224} {125} (1973).

\bibitem{gibb05} G. W. Gibbons et al., \textit{Phys. Rev.} {\bf D72}, {084028} (2005).

\bibitem{bray} H. L. Bray, \textit{ Notices of the AMS} {Vol. 49} {No. 11}, {1372} (2002).

\bibitem{bray1} H. L. Bray and P. T. Chru\'{s}ciel, \textit{arXiv: gr-qc/0312047}, (2004).

\bibitem{jang} P. S. Jang and R. M. Wald, \textit{J. Math. Phys. } {\bf 18}, {41} (1977).

\bibitem{rg} R. Geroch,  \textit{ Ann. N. Y. Acad. Sci. } {\bf 224}, {108} (1973).

\bibitem{gibb99} G. W. Gibbons, \textit{ Class. Quant. Grav.} {\bf 16}, {1677} (1999). 

\bibitem{yau} R. Schoen and S. T. Yau, \textit{Comm. Math.  Phys.} {\bf 65}, {45} (1979).  

\bibitem{hui} G. Huisken and T. Ilamanen, \textit{J. Diff. Geom.} {59} (353) (2001).

\bibitem{lmp09} H. Lu et al., \textit{Phys. Rev. Lett. } {\bf 103}, {091301} (2009).

\bibitem{ym09} Y. S. Myung, \textit{Phys. Lett. } {B 678}, {127} (2009).

\bibitem{mk10} Y. S. Myung and Y. W. Kim, \textit{Euro. Phys. J. C} {68}, {265} (2010).

\bibitem{cc9a} R. G. Cai et al., \textit{Phys. Rev. D.} {\bf 80}, {024003} (2009).

\bibitem{cc9b} R. G. Cai et al., \textit{Phys. Lett.} {\bf B 679} {504} (2009).


\bibitem{xu}  W.~Xu et al., \textit{Phys. Lett. } {B 746} {53} (2015).

\bibitem{ls73} L. Smarr, \textit{Phys. Rev. Lett.} {\bf 30} 71 (1973).

\bibitem{bcw73} J. M. Bardeen et al., {\it Commun. Math. Phys.} {\bf 31}, {161} (1973).

\bibitem{cd70} D. Christodoulou, \textit{Phys. Rev. Lett.} {\bf 25} {1596} (1970).


\bibitem{cv96} M. Cveti\v{c} and D. Youm, 
\textit{ Phys. Rev. }{\bf D 54} {2612} (1996).

\bibitem{cv97} M. Cveti\v{c} and F. Larsen, 
\textit{ Nucl. Rhys.} {\bf B 506} {107} (1997).

\bibitem{cvf97} M. Cveti\v{c} and F. Larsen, 
\textit{ Phys. Rev.} {\bf D 56} {4994} (1997).

\bibitem{guica} M. Guica et al., 
\textit{ Phys. Rev.} {\bf D 80}, {124008} (2009).

\bibitem{brena} W. G. Brenna et al., \textit{Phys. Rev.} {\bf D 92} {044015} (2015).

\bibitem{ac10} A. Castro et al., \textit{ Phys. Rev.} {\bf D 82} {024008} (2010).

\bibitem{ok92} I. Okamoto and O. Kaburki, \textit{ Mon. Not. R. Astr. Soc.}  {\bf 255} {539} (1992).


\end{thebibliography}
\end{document}